\documentclass[12pt,preprint]{aastex}
\usepackage{emulateapj5}
\usepackage{epsfig}
\usepackage{natbib}

\newcommand{\etal}{et al.}

\newcommand{\flamb}{erg s$^{-1}$ cm$^{-2}$ \AA$^{-1}$}

\newcommand{\feii}{\ion{Fe}{2}}

\def\gtrsim{\mathrel{\hbox{\rlap{\hbox{\lower4pt\hbox{$\sim$}}}\hbox{\raise2pt\hbox{$>$}}}}}
\newcommand{\fwha}{\ensuremath{\mathrm{FWHM}_\mathrm{H{\alpha}}}}
\newcommand{\fwhb}{\ensuremath{\mathrm{FWHM}_\mathrm{H{\beta}}}}
\newcommand{\halpha}{H\ensuremath{\alpha}}

\newcommand{\hbeta}{H\ensuremath{\beta}}

\newcommand{\kms}{km~s\ensuremath{^{-1}}}

\newcommand{\lf}{\ensuremath{L_{5100}}}

\newcommand{\lha}{\ensuremath{L_{\mathrm{H{\alpha}}}}}
\newcommand{\lhb}{\ensuremath{L_{\mathrm{H{\beta}}}}}

\newcommand{\mbh}{\ensuremath{M_\mathrm{BH}}}

\newcommand{\msigma}{\ensuremath{M_{\mathrm{BH}}-\sigmastar}}
\newcommand{\msun}{\ensuremath{M_{\odot}}}

\newcommand{\nii}{[\ion{N}{2}]}

\newcommand{\oiii}{[\ion{O}{3}]}

\newcommand{\rblr}{\ensuremath{R_{\mathrm{BLR}}}}

\newcommand{\sii}{[\ion{S}{2}]}
\newcommand{\sigmastar}{\ensuremath{\sigma_{\ast}}}

\def\lax{{$\mathrel{\hbox{\rlap{\hbox{\lower4pt\hbox{$\sim$}}}\hbox{$<$}}}$}}
\def\gax{{$\mathrel{\hbox{\rlap{\hbox{\lower4pt\hbox{$\sim$}}}\hbox{$>$}}}$}}

\slugcomment{To appear in {\it The Astrophysical Journal}.}
\shorttitle{Black Hole Masses in AGNs from \halpha}
\shortauthors{GREENE \& HO}

\received{2005 May 06}
\begin{document}

\title{Estimating Black Hole Masses in Active Galaxies Using the
\halpha\ Emission Line}

\author{Jenny E. Greene}
\affil{Harvard-Smithsonian Center for Astrophysics, 60 Garden St., 
Cambridge, MA 02138}

\and

\author{Luis C. Ho}
\affil{The Observatories of the Carnegie Institution of Washington,
813 Santa Barbara St., Pasadena, CA 91101}

\begin{abstract}
It has been established that virial masses for black holes in
low-redshift active galaxies can be estimated from measurements of the
optical continuum strength and the width of the broad \hbeta\ line.
Under various circumstances, however, both of these quantities can be
challenging to measure or can be subject to large systematic
uncertainties.  To mitigate these difficulties, we present a new method
for estimating black hole masses.  From analysis of a new sample of
broad-line active galactic nuclei, we find that \halpha\ luminosity
scales almost linearly with optical continuum luminosity and that a
strong correlation exists between \halpha\ and \hbeta\ line widths.
These two empirical correlations allow us to translate the standard
virial mass system to a new one based solely on observations of the
broad \halpha\ emission line.
\end{abstract}

\keywords{galaxies: active --- galaxies: jets --- galaxies: nuclei --- 
galaxies: Seyfert --- quasars: general}

\section{Introduction}

The recent discovery of a tight correlation between black hole (BH)
mass and the stellar velocity dispersion of the bulge of the host
galaxy (the \msigma\ relation; Gebhardt \etal\ 2000a; Ferrarese \&
Merritt 2000) has sparked renewed interest in deriving BH masses in
active galactic nuclei (AGNs).  Reverberation mapping (e.g.,~Blandford
\& McKee 1982) experiments can yield estimates of the radius of the
broad-line region (BLR) from the lag between the variability in the
AGN continuum and the corresponding variability in the broad permitted
lines.  With the BLR radius (\rblr) in hand and some measure of the
BLR velocity dispersion ($v$), one can infer a virial mass for the BH,
\mbh~$=v^2$\rblr/$G$ (Ho 1999; Wandel, Peterson, \& Malkan 1999; Kaspi
et al. 2000).  Remarkably, this simple prescription yields BH masses
that seem to be accurate to a factor of $\sim$3 when compared against
the \msigma\ relation (Gebhardt et al. 2000b; Ferrarese et al.  2001;
Nelson et al. 2004; Onken et al. 2004).  The number of AGNs with BH
masses derived from reverberation mapping is currently small ($35$), 
but \mbh\ can be estimated indirectly from single-epoch spectra
using an empirical correlation between \rblr\ and optical continuum
luminosity that has been calibrated using the reverberation-mapped
sample: \rblr\ $\propto L_{5100}^{0.7}$ (Kaspi \etal\ 2000), where
\lf\ = $\lambda L_{\lambda}$ at $\lambda$ = 5100 \AA.  From a
single-epoch spectrum one can measure both $v$ and \lf, and thereby
infer \mbh.

While virial estimates of BH masses have been widely used in the
literature (e.g., Vestergaard 2002; Greene \& Ho 2004), various
systematic uncertainties may affect the measurements of both
luminosity and velocity.  
Broad \feii\ emission in Type 1 AGNs can add ambiguity in the 
determination of the optical continuum luminosity at 5100 \AA.  
In lower-luminosity AGNs, dilution by host
galaxy starlight can be so severe that it becomes very difficult, if
not virtually impossible, to isolate the nonstellar continuum
unambiguously.  As noted by Wu \etal\ (2004), an additional
complication arises in sources, such as blazars, whose optical
continuum can be subject to contamination by synchrotron emission
from jets, which might be partly beamed.  Under these circumstances,
the observed \lf\ overestimates the true thermal component that
governs the BLR size-luminosity relation.

For nearby AGNs, including the primary reverberation-mapped sample
(Kaspi et al. 2000), the velocity dispersion used in the virial
expression is usually measured through the width of the broad \hbeta\
line.  However, \hbeta\ is at least a factor of 3 weaker than \halpha,
and so from considerations of signal-to-noise ratio (S/N) alone
\halpha, if available, is superior to \hbeta.  In practice, in some
cases \halpha\ may be the {\it only}\ detectable broad permitted line
in the optical (traditionally such objects are known as Seyfert 1.9
galaxies; Osterbrock 1981).

In an effort to minimize uncertainties in virial BH mass estimates,
and to broaden their applicability to classes of AGNs that would
otherwise be inaccessible using the conventional methodology (e.g.,
blazars, galaxy-dominated low-luminosity sources, Type 1.9 Seyferts),
we propose to translate the virial mass system of Kaspi et al. (2000),
which makes use of the 5100 \AA\ continuum luminosity and the broad
\hbeta\ line width, to a new system based {\it entirely}\ on
observations of the broad \halpha\ line.  We achieve this in two
steps.  First, we revisit the well-known empirical correlation between
optical continuum luminosity and Balmer emission-line luminosity (Yee
1980; Shuder 1981; Ho \& Peng 2001), which now can be established with
much greater precision using large, homogeneous samples of AGNs.  This
allows us to replace \lf\ with \lha, which has the advantage of being
relatively straightforward to measure even for low-luminosity or
galaxy-dominated sources (e.g., Ho et al. 1997; Greene \& Ho 2004).
Unlike the optical continuum, the line emission is also expected to be
little affected by the jet component in radio-loud sources or blazars
(e.g., Wang, Ho, \& Staubert 2003).  This approach is similar to that
taken by Wu et al.  (2004), who converted \lf\ to \lhb, but their
analysis was confined to the original reverberation mapping sample of
Kaspi et al. (2000), which consists, by necessity, of AGNs that are variable 
on timescales probed by current reverberation mapping experiments.  This 
selection effect may result in subtle biases, and, in any case, the sample 
suffers from limited statistics.  Secondly, we substitute
the line width of \hbeta\ with that of \halpha, using a newly derived
empirical correlation between the two.

Throughout we assume the following cosmological parameters to calculate
distances: $H_0 = 100h = 71$~\kms~Mpc$^{-1}$, $\Omega_{\rm m} = 0.27$,
and $\Omega_{\Lambda} = 0.75$ (Spergel \etal\ 2003).

\begin{figure*}[t]
\psfig{file=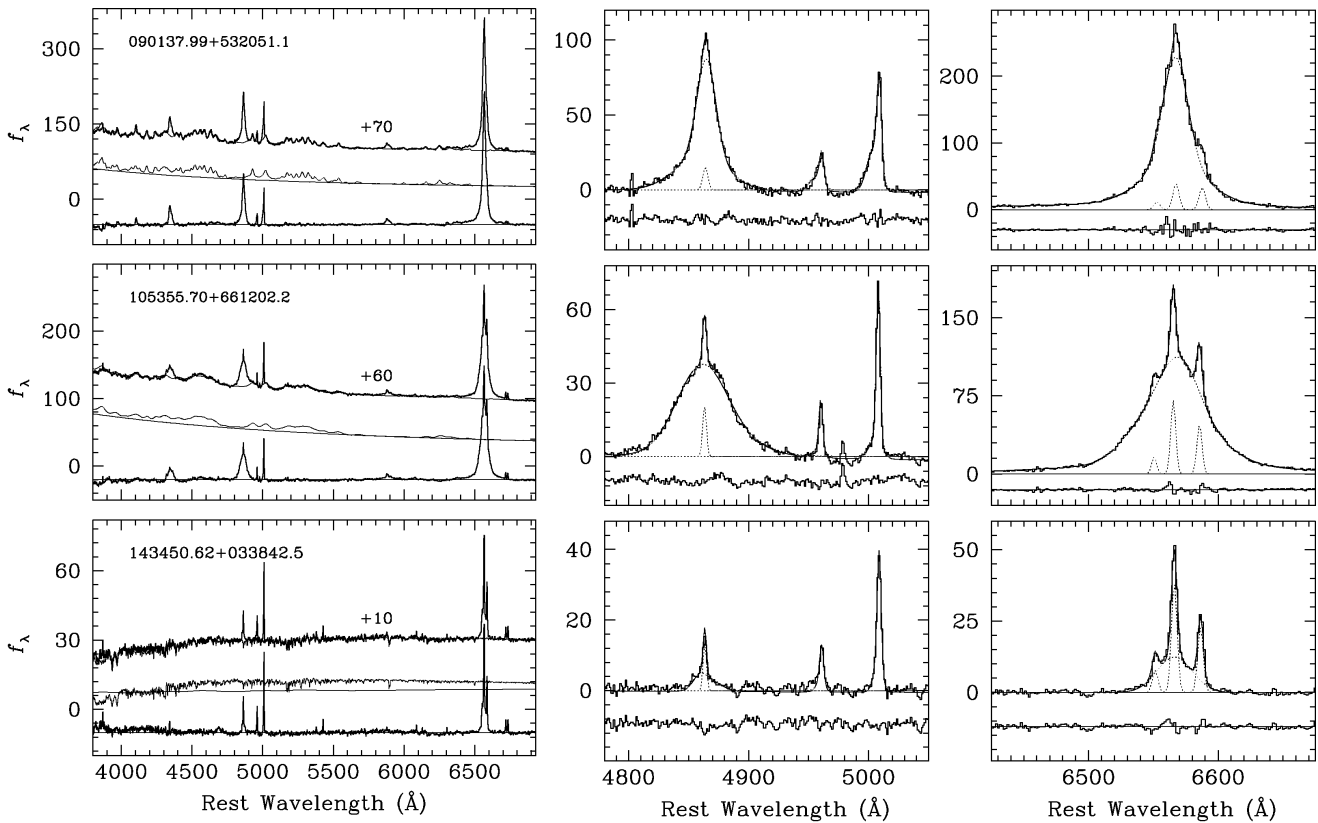,width=0.51\textwidth,angle=-90}
\vskip +5mm \figcaption[]{ 
Sample spectra for three objects in our sample.  The
ordinate is in units of 10$^{-17}$ \flamb.  {\it Left}: In each panel,
the top shows the observed spectrum overplotted with the model for the
continuum, offset in the ordinate for clarity.  The middle shows the
fitted components (power law, Fe {\tiny II} template, and, if
necessary, host galaxy) of the continuum model, with no offset.  The
bottom shows the continuum-subtracted, emission-line spectrum.  {\it
Center}: Fits to the \hbeta+[O{\tiny III}]~$\lambda\lambda 4959,~5007$
region.  The data are shown in histogram, the fitted broad and narrow
components are shown in dotted lines, and the final model (sum of all
components) is shown as a solid line.  The residuals are plotted
below.  {\it Right}: Same as the middle panels, but for the \halpha+[N
{\tiny II}]~$\lambda\lambda 6548,~6583$ complex.
}
\end{figure*}

\section{The Sample}

We utilize the large and homogeneous database provided by the Sloan
Digital Sky Survey (SDSS; York \etal\ 2000), which will eventually
obtain imaging and follow-up spectroscopy for about one-quarter of the
sky.  Briefly, spectroscopic candidates are selected based on
multi-band imaging (Strauss \etal\ 2002) with a drift-scan camera
(Gunn \etal\ 1998).  A pair of spectrographs is fed by
$3\arcsec$-diameter fibers, covering $\sim 3800 - 9200$ \AA\ with an
instrumental resolution of $\lambda/\Delta \lambda \approx 1800$
(Gaussian $\sigma_{\rm inst} \approx 71$ \kms).  Integration times are
determined for a minimum S/N = 4 at $g=20.2$ mag.  The spectroscopic
pipeline performs basic image calibrations, as well as spectral
extraction, sky subtraction, removal of atmospheric absorption bands,
and wavelength and spectrophotometric calibration (Stoughton \etal\
2002).  Finally, spectral classification is performed using
cross-calibration with stellar, emission-line, and active galaxy
templates. The Third Data Release (DR3; Abazajian \etal\ 2005) of SDSS
contains 45,260 spectroscopically identified AGNs with $z < 2.3$.  Of
these, $\sim 3000$ have $z \leq 0.35$, and hence \halpha\ in the
bandpass.

From this sample we select objects with a high S/N ($\geq 25$ per
pixel) and low galaxy contamination, so that we can unambiguously
measure \lf.  We use the equivalent width (EW) of the Ca~{\sc ii}
K~$\lambda 3934$ line to estimate the degree of galaxy contamination;
within the SDSS bandpass this is the highest-EW stellar absorption
feature that is not confused with line emission.  By artificially
diluting composite galaxy spectra\footnote{The composite spectra are 
constructed from a library of $\sim 9000$  early-type galaxies 
from SDSS, and span a wide range in velocity dispersion, redshift, 
luminosity, and effective radius.  We degrade the S/N ratio of the
composites to match that of our sample.} from Bernardi \etal\ (2003)
with varying levels of featureless AGN
continuum (which for a narrow spectral region is well approximated by
a constant), we find that an EW cutoff of 1.5 \AA\ for the Ca~{\sc
ii}~K line corresponds to a galaxy contribution of $\leq 20 \%$.
Thus, our primary sample was chosen to satisfy EW(Ca~K) $\leq 1.5$
\AA.  In order to increase the dynamic range of the sample at the
lowest luminosities, we included $\sim 30$ more galaxies, including
the 19 objects from Greene \& Ho (2004), which have a mean S/N $\approx$ 17 
and EW(Ca~K) as high as 10 \AA.  The total sample consists of 229
objects.

To investigate the possible influence of jet emission on \lf, we also
consider an auxiliary sample of 59 radio-loud objects, selected using
the standard criterion for radio-loudness, $R \geq 10$, where $R$ is
the ratio of the flux densities at 6~cm and 4400 \AA\ (Kellermann et
al. 1989).  The radio flux densities were obtained from the Faint
Images of the Radio Sky at Twenty-cm (FIRST; Becker, White, \& Helfand
1995) survey, which covers all the objects in our sample.  We
corrected the flux densities from 20~cm to 6 cm assuming a spectrum
$f_{\nu} \propto \nu^{-0.46}$ (Ho \& Ulvestad 2001).  The derived $R$
values for the radio-loud sample range from 10 to $10^4$.

\section{Spectral Fitting}

Below we discuss our method to measure the AGN continuum and the broad
emission lines.  The continuum and line measurements are corrected for
Galactic extinction using the extinction map of Schlegel, Finkbeiner,
\& Davis (1998) and the reddening law of Cardelli, Clayton, \& Mathis
(1989).

\subsection{Continuum}

The nonstellar, featureless continuum of Type 1 (broad-line) AGNs is
challenging to measure at optical wavelengths because it is entangled
with a complex emission-line spectrum and starlight, depending on the
relative brightness of the nucleus with respect to the host galaxy.
Particularly troublesome are the broad, blended \feii\ multiplets,
which form a ``pseudo-continuum'' that litters most of the spectrum,
but especially the regions flanking \hbeta.  Moreover, the intrinsic
shape of the featureless continuum itself is not well known.  For the
objects that have minimal starlight contamination, which, by
selection, constitute the majority of our sample, we decompose the
spectrum by simultaneously fitting a two-component model consisting of
(1) an underlying featureless continuum and (2) an empirical \feii\
template.  In practice, the featureless continuum over the optical
region can be approximated well by a double power law, with a spectral
break at $\lambda \approx 5000$ \AA\ (Vanden Berk \etal\ 2001; M. Kim
\& L. C. Ho 2005, in preparation).  We require that the combined flux of the 
two power-law components at 5600 \AA, a relatively line-free region, be equal
to the observed flux at that point.
Following Boroson \& Green (1992), an effective \feii\ template can be
generated by a simple velocity-broadening and scaling of the \feii\
spectrum derived from observations of the ``narrow-line'' Type~1 AGN
{\sc I}~Zw~1 (kindly provided by T. Boroson).  In order to avoid
contamination from strong spectral lines, we only include the
following regions in our fit: 4170--4260, 4430--4770, 5080--5550,
6050--6200, and 6890--7010 \AA.  We do not fit the region below 3685
\AA, since it is not covered by the {\sc I}~Zw~1 \feii\ template of
Boroson \& Green (1992); consequently, we do not have to model
the Balmer continuum (e.g., Wills, Netzer, \& Wills 1985).

Approximately 30 objects at the lowest luminosities have sufficient
galaxy contamination to warrant removing it.  In lieu of more
sophisticated methods for starlight decomposition (e.g., Greene \& Ho
2004, 2005), we find that for the present sample the galaxy continuum
can be modeled adequately using a scaled spectrum of a
velocity-broadened K giant star (obtained from the SDSS).

\subsection{Emission Lines}

The primary challenge in measuring accurate fluxes and profiles of
broad emission lines is to properly deblend contaminating narrow-line
components.  Here we are only concerned with the region immediately
surrounding \halpha\ and \hbeta.  For \halpha, as outlined in Greene
\& Ho (2004; see also Ho et al. 1997), we begin by constructing a
model for the narrow components using the \sii~$\lambda \lambda$6716,
6731 lines.  In short, a multi-Gaussian model is fit to the \sii\
doublet, and then shifted and scaled to fit the
\halpha+\nii~$\lambda\lambda$6548, 6583 complex.  The relative
positions of the narrow \halpha\ and \nii\ lines are constrained by
their laboratory values, and the relative ratio of the two \nii\
components is fixed to 2.96.  We then fit the broad component of
\halpha\ with as many Gaussian components as needed to provide a
statistically acceptable fit.  This method provides the needed
flexibility to reproduce highly non-Gaussian, asymmetric profiles, but
we attach no physical significance to the individual components.

The fitting procedure for \hbeta\ is similar to that outlined above;
in this case the narrow-line contamination comes from the narrow
component of \hbeta\ and \oiii~$\lambda\lambda$4959, 5007.  Because of
their proximity, we first experimented with modeling the narrow
\hbeta\ component using \oiii\ as a template, but could not reliably
obtain a satisfactory fit. The mismatch stems partly from the
frequently observed asymmetric blue wing of \oiii\ (Greene \& Ho
2005)\footnote{ The blue asymmetric wing of [O {\tiny III}] is
presumably absent from the narrow component of \halpha\ and \hbeta,
since [S~{\tiny II}], which generally lacks the asymmetric component
(Greene \& Ho 2005), fits \halpha\ well.}.  The profile of \oiii\ is
also expected to be a poor representation of the profile of \sii\ (and
hence narrow \halpha\ and \hbeta) if the narrow-line region is
stratified in density (see discussion in Greene \& Ho 2005).
Therefore, as with \halpha, we use the template built from \sii\ to
model narrow \hbeta.  The \hbeta\ centroid is fixed relative to the
narrow \halpha\ position while its flux is limited by the value
appropriate for Case B$\arcmin$ recombination (\halpha\ = 3.1 \hbeta;
Osterbrock 1989).  We simultaneously fit \oiii\ using a two-component
Gaussian (identical for both lines of the doublet), one for the core
and another for the blue wing, as described in Greene \& Ho (2005).
Finally, the broad \hbeta\ profile is modeled as a multi-component
Gaussian, analogous to \halpha.  While in general we find that this
procedure provides an acceptable fit to the entire region, there were
some cases ($\sim 20$) for which we could not obtain an acceptable fit
for the narrow \hbeta\ using the \sii\ model; this may be due to gross
variations in spectral resolution between the blue and red ends of the
spectrum.  In these cases, in order to fit the narrow \hbeta\
component, we retain the flux and centroid limits described above but
relax the line width constraint, and instead impose an (rather
arbitrary) upper limit of $\sigma \leq 550$ \kms.

\begin{figure*}
\begin{center}
\vskip -1.truein
\epsfig{file=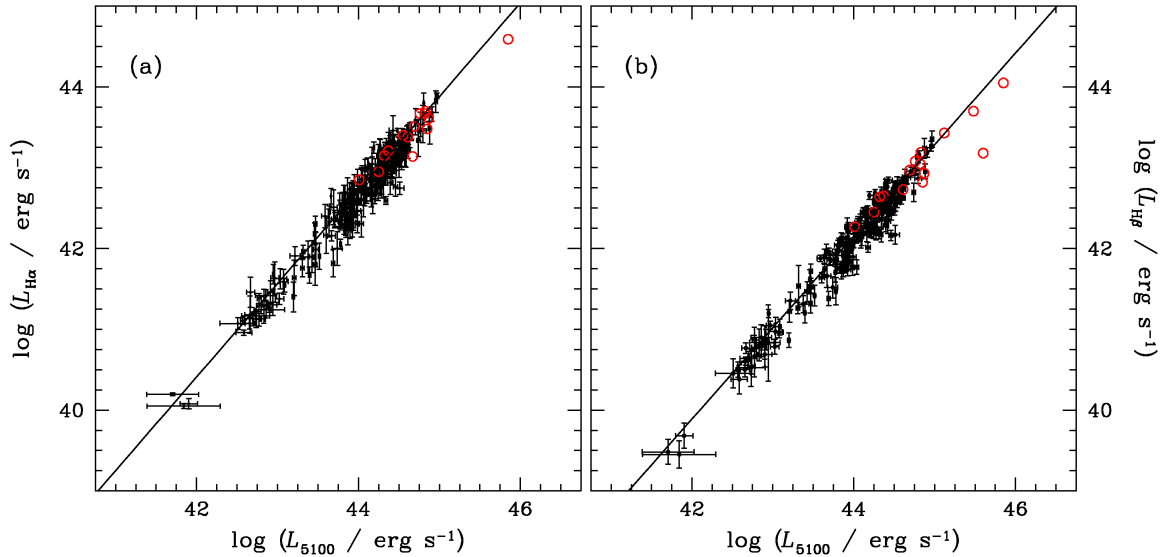,width=0.65\textwidth,keepaspectratio=true,angle=-90}
\vskip -8mm
\figcaption[]{
The correlation between \lf\ and ({\it a}) \lha\ and ({\it b}) \lhb\
for the combined broad and narrow components of the line. Filled points are 
our measurements, and open (red) points are from Kaspi \etal\ (2000).
The error bars for the points from Kaspi \etal\ are smaller than the symbol
size.  The line gives the ordinary least-squares fit to the data.
\label{exspecfig}}
\vskip -2mm
\end{center}
\end{figure*}

\subsection{Errors and Uncertainties}

In objects with minimal galaxy contamination, the uncertainty in the
flux of the optical continuum is dominated by the proper removal of
\feii\ emission and errors in the continuum slope.  We use the
measured \feii\ flux at 5100 \AA\ to estimate the uncertainty
contributed by our decomposition.  To estimate errors resulting from
the power-law fit, we average the fluxes obtained when the slope of
the power law is allowed to increase or decrease by one standard
deviation, as obtained from the fit.  The final adopted error is the
quadrature sum of the \feii\ and slope errors, although the latter
usually dominates.  Typical errors are $\sim 0.04$ dex.  The largest
source of uncertainty in measurements of line flux comes from the
placement of the continuum level.  We estimate our uncertainty by
adjusting the continuum level by one standard deviation above and
below the nominal best-fit level of the continuum, and then
recalculating the line flux.  Typical errors in the flux are $\sim 35
\%$, or 0.13 dex.

Estimating errors in line width is more difficult, since uncertainty
in the decomposition between broad and narrow components dominates the
measurement error.  In some cases the separation is obvious, but in
others (see, e.g., the top panel of Fig. 1) there is no clear
distinction between the two components.  We generate an artificial
spectrum for each object by combining the best-fit model emission
lines and Gaussian random errors with matching S/N.  An artificial
spectrum generated in this manner will suffer from the same
decomposition ambiguity as the original data.  We then fit these
artificial spectra using our fitting procedure, as outlined above.
The estimated uncertainty is simply taken to be the difference between
the model and measured FWHM.  We find that S/N significantly impacts
the reliability of our fits.  While typical errors are $\sim 7 \%$ (or
0.03 dex) and $\sim 10 \%$ (0.05 dex) for \fwha\ and \fwhb,
respectively, these are found to increase to as much as 0.1 dex in
cases with the lowest S/N.

\section{Results}

\subsection{Empirical Correlations}

Our data reveal a well-defined correlation between Balmer emission-line 
luminosity and \lf.  This is illustrated in Figure 2, which includes 
the reverberation-mapped sample from Kaspi \etal\ (2000).  
Treating \lf\ as the independent variable (under the assumption that \lf\
traces the continuum that powers the line emission) and accounting for errors
in both variables (Press \etal\ 1992), an ordinary least-squares fit yields

\begin{equation}
\lha = (5.25 \pm 0.02) \times 10^{42} 
\left( \frac{\lf}{10^{44}~{\rm erg~s^{-1}}}\right)^{(1.157 \pm
                          0.005)}~{\rm erg~s^{-1}} 
\end{equation}

\noindent
and

\begin{equation}
\lhb = (1.425 \pm 0.007) \times 10^{42} 
\left(\frac{\lf}{10^{44}~{\rm erg~
s^{-1}}} \right)^{(1.133 \pm 0.005)}~{\rm erg~s^{-1}}.
\end{equation}
\vskip +0.35cm

\noindent 
The rms scatter in the above luminosity-luminosity relations is quite
small, being only $\sim 0.2$ dex in each case.  The correlation
strengths are correspondingly strong, with Kendall's $\tau = 0.8$, and
do not improve when FWHM is 
included as a third parameter using the
partial correlation test of Akritas \& Siebert (1996).  These
relations are fit for the combined broad and narrow components.
However, the best-fit parameters are virtually unchanged when only the
broad component is considered, since the narrow component accounts for
only $\sim 7 \%$ and $\sim 10 \%$ of the total \halpha\ and \hbeta\
flux, respectively.  Note that the slopes of both correlations are
formally larger than unity, which can be excluded at a confidence
level of $>99$\% based on bootstrap simulations.

\epsfig{file=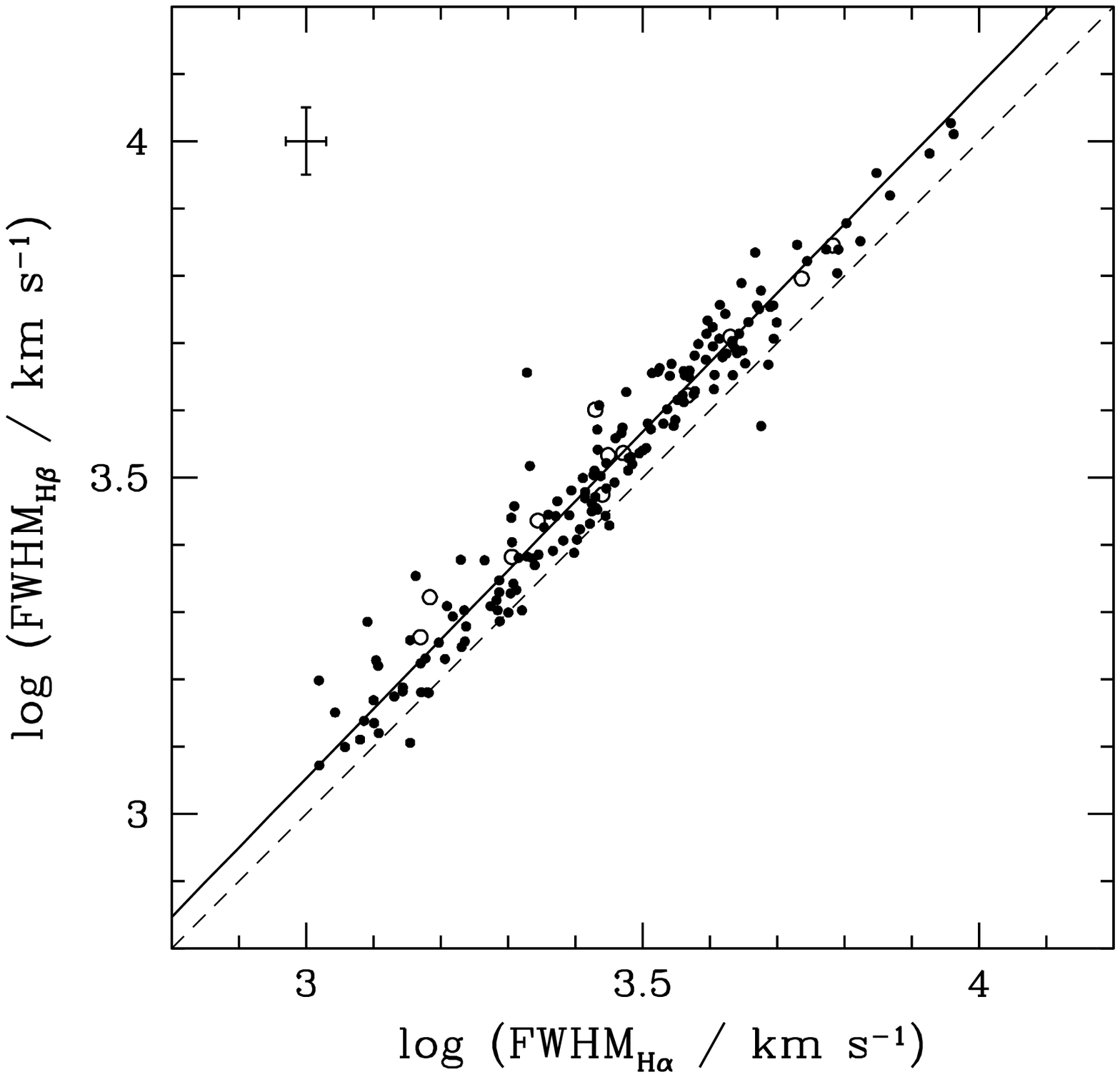,width=0.45\textwidth,keepaspectratio=true}
\vskip -1mm \figcaption[]{ 
The correlation between \halpha\ and
\hbeta\ line widths.  Filled points are our measurements, and open 
points are from Kaspi \etal\ (2000).  The solid line gives the
ordinary least-squares fit to the 162 objects with EW(\hbeta) $> 50$
\AA.  The dashed line denotes \fwha\ = \fwhb.  A typical error bar is
shown in the top-left corner.
\label{hahbfwhm}}
\vskip +5mm

We also find a well-defined correlation between the \halpha\ and
\hbeta\ line width (Fig. 3), which we parameterize\footnote{Peterson
et al.  (2004) advocate the use of the actual line dispersion (second
moment of the line profile).  We have adopted the FWHM here for
consistency with the formalism of Kaspi et al. (2000).  We also find
that the empirical correlation between the line widths of \halpha\ and
\hbeta\ is significantly tighter using FWHM than the line dispersion.}
by the FWHM.  In this case, neither \fwha\ nor \fwhb\ can be regarded
as the independent variable, so we calculate the ordinary
least-squares bisector, again accounting for errors in both parameters
(Akritas \& Bershady 1996).  Whereas for the line-continuum
correlations we were forced to include lower-S/N spectra in order to
increase the dynamic range in luminosity, in this case we can span the
entire range in line width without considering the full sample.  To
minimize errors resulting from profile decomposition, we concentrate
on spectra with strong \hbeta\ lines by excluding objects with \hbeta\
EWs less than 50 \AA.  For this restricted sample of 162 objects, we
find

\begin{equation}
\fwhb = (1.07 \pm 0.07) \times 10^3 
\left(\frac{\fwha}{10^3~\mathrm{\kms}}\right)^{(1.03 \pm
0.03)}~\mathrm{\kms}.
\end{equation}
\vskip +0.35cm

\noindent
The rms scatter around the best-fit line is $\sim 0.1$ dex.
We note that when we include the entire sample, or look at the radio-loud
subset separately, \hbeta\ is still on average broader than \halpha, but the 
scatter is larger and the slope 
of the correlation changes slightly.

\subsection{Radio-loud Sources}

If the optical continuum of radio-loud sources is significantly
boosted by nonthermal emission from a jet, their \lf, and hence virial
mass, will be systematically overestimated.  We can empirically test
the magnitude of this effect by examining the correlation between \lf\
and Balmer-line luminosity for our auxiliary radio-loud sample.  If
jet contamination is important, we would expect the 
\epsfig{file=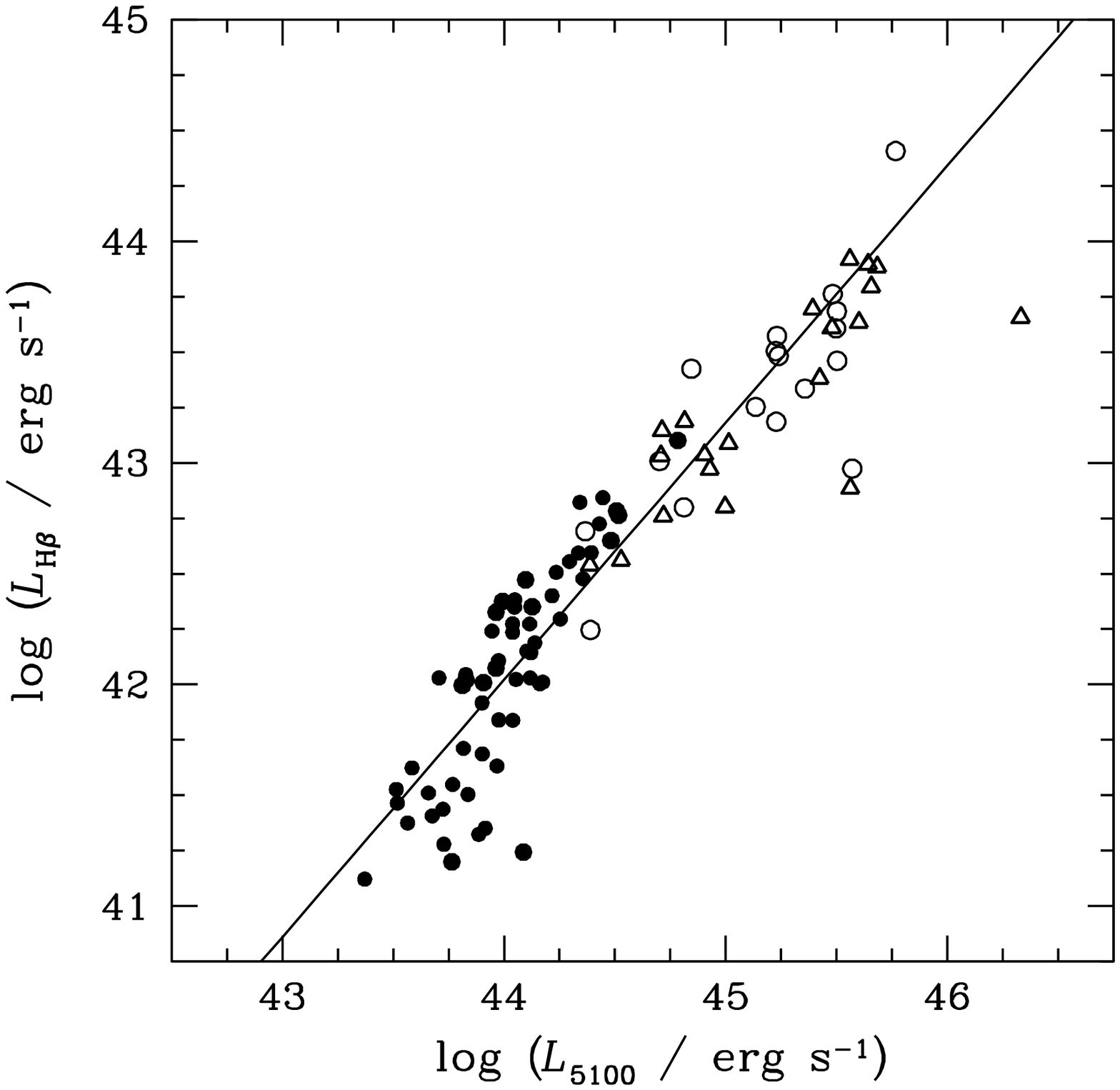,width=0.45\textwidth,keepaspectratio=true}
\vskip -1mm
\figcaption[]{
The correlation between \lf\ and \lhb\
luminosity (broad + narrow components) for the radio-loud sources.
Filled symbols are from our sample; open symbols are from Brotherton
(1996).  Core-dominated sources with core-to-lobe luminosity ratios
$\geq 1$ are marked as triangles.  The line represents our best fit to
the \lhb-\lf\ relation shown in Figure 2{\it b}.
\label{l51hbrl}}
\vskip +5mm
\noindent
radio-loud objects
to lie systematically to the right of the best-fit relations shown in
Figure 2.  However, as shown in Figure 4, the radio-loud sources
(filled points) do not show a significant offset with respect to the
main sample.  Their mean displacement from our best-fit \lhb-\lf\ line
is only +0.02 dex in \lf, and for the strongest radio sources ($R \geq
100$) the mean displacement is in fact $-$0.04; neither result is
statistically significant.  Instead of the degree of radio-loudness,
the more important parameter may be the amount of beaming, which can
be guaged roughly by the degree of core dominance of the radio
emission.  Unfortunately, the resolution of FIRST (5\arcsec\ or 10 kpc
at $z=0.1$) is inadequate to distinguish truly core-dominated objects
from objects with intrinsically compact jets.  We thus additionally
consider objects from the sample of Brotherton (1996), which is
composed entirely of radio-loud objects with tabulated core-to-lobe
radio luminosity ratios.  For the 37 Brotherton objects with measured
$V$-band magnitudes and $R$ values, we calculate \lf\ assuming an
intrinsic AGN continuum spectrum of $f_{\nu} \propto \nu^{-0.45}$
(Vanden Berk \etal\ 2001).  Brotherton's sample is, on average, about
an order of magnitude more luminous than ours.  The entire Brotherton
sample (open symbols in Fig. 4) 
is displaced by +0.1 dex from the
fiducial line of radio-quiet objects, increasing to +0.14 dex for the
core-dominated systems (defined as those with core-to-lobe ratios
$\geq 1$; open triangles).  We conclude that BH masses based on \lf\
can be systematically overestimated in radio-loud objects, but the
effect is only serious for relatively luminous, core-dominated
sources.  While the systematic enhancement of the optical continuum
relative to Balmer line emission in the most radio-loud, core-dominant
sources can be attributed reasonably to jet contamination, we cannot
rule out the possibility that this effect is due to changes in the
ionizing spectrum or covering factor of the line-emitting gas.

\subsection{A New Formalism for Estimating BH Virial Masses}

The existence of the tight empirical correlations described above
(Eqs. 1 and 3) implies that we can transform the virial mass
expression of Kaspi et al.  (2000), which makes use of \lf\ and \fwhb,
into a new formalism that depends solely on the observed properties of
the broad \halpha\ line (namely \lha\ and \fwha, derived from a
single-epoch spectrum), without incurring appreciable additional
uncertainty.

We begin by rederiving the radius-luminosity relation of Kaspi et
al. (2000), using the more recent \hbeta\ lags (derived from the
cross-correlation centroid) and \lf\ luminosities listed in Peterson
et al. (2004), corrected to our adopted cosmology.  The data points
that Peterson et al. deem to be uncertain are omitted.  We calculate
symmetric error bars using the mean of the quoted upper and lower
uncertainties, and use the weighted average of all available lag
measurements.  Using an ordinary least-squares fit with uncertainties
in both parameters, we obtain

\begin{equation}
\rblr = (30.2 \pm 1.4)~\left( \frac{\lf}{10^{44}~{\rm erg~
s^{-1}}} \right)^{0.64 \pm 0.02} \ \ \ {\rm lt-days} .
\end{equation}

\vskip +0.5cm

\noindent
Note that the slope of the radius-luminosity relation, $0.64 \pm
0.02$, is somewhat shallower than that originally published by Kaspi
et al. (2000), $0.700 \pm 0.033$, although it is consistent with the
value of $0.67 \pm 0.07$ recently rederived by Kaspi \etal\ (2005).
For $v = \frac{\sqrt{3}}{2}~v_{\mathrm{FWHM}}$, appropriate for random
orbits, and using $v_{\mathrm{FWHM}}$ = \fwhb\ for consistency with
Kaspi \etal\ (2000), the virial formula\footnote{By comparing
reverberation mapping masses with an enlarged sample of bulge velocity
dispersions, Onken et al. (2004) advocate a normalization factor for
the virial formula larger than the one we adopted by a factor of 1.8.}
for BH mass is

\begin{eqnarray}
\mbh &=& (4.4 \pm 0.2) \times 10^6 \nonumber \\
     & & \hskip -0.3in \left(
\frac{\lf}{10^{44}~{\rm erg~
s^{-1}}} \right)^{0.64 \pm 0.02}~\left( \frac{\fwhb}{10^{3}~\mathrm{km~s^{-1}}} \right)^2 \msun .
\end{eqnarray}
\vskip +0.5cm

\noindent
Substituting Equations 1 and 3 into Equation 5, we obtain a virial mass 
formula that depends on the \halpha\ line alone:

\begin{eqnarray}
\mbh & = & (2.0 ^{+0.4}_{-0.3}) \times 10^6 \nonumber \\
     &   & \hskip -0.5in \left(
\frac{\lha}{10^{42}~{\rm erg~s^{-1}}} \right)^{0.55 \pm 0.02}\left(
\frac{\fwha}{10^{3}~\mathrm{\kms}} \right)^{2.06 \pm 0.06} \msun .
\end{eqnarray}
\vskip +0.5cm

While we argue that \halpha\ is the line of choice whenever possible, we 
recognize that often \hbeta\ is more readily available (for instance, due to
redshift constraints).  In such cases, it might still be preferable to use
\lhb\ rather than \lf, and for completeness we give the virial mass 
formula based only on the \hbeta\ line:

\begin{eqnarray}
\mbh &=& (3.6 \pm 0.2) \times 10^6 \nonumber \\
     & & \hskip -0.3in \left(
\frac{\lhb}{10^{42}~{\rm erg~s^{-1}}} \right)^{0.56 \pm 0.02}~\left(
\frac{\fwhb}{10^{3}~\mathrm{\kms}} \right)^2 \msun .
\end{eqnarray}
\vskip +0.5cm

\subsection{Comparison with Kaspi et al. (2005)}

Kaspi \etal\ (2005) revisit the radius-luminosity relation for the 
reverberation-mapped sample, and we present a brief comparison of 
our respective results.  Our best-fit \rblr-\lf\ slope, 0.64 $\pm$ 0.02, 
is consistent with the result of Kaspi \etal, who find a best-fit slope
of 0.67 $\pm$ 0.07 for \rblr\ derived from \hbeta\ lags alone (line 18 of 
their Table 3).  The slight discrepancy between our value and theirs stems from
a combination of the slightly differing cosmologies, samples, and fitting 
techniques employed in the two studies.  For completeness, we recast our BH 
mass estimators based on the radius-luminosity relation of Kaspi \etal\ (2005):

\begin{equation}
\rblr = (20.0^{+2.8}_{-2.4})~\left( \frac{\lf}{10^{44}~{\rm erg~
s^{-1}}} \right)^{0.67 \pm 0.07} \ \ \ {\rm lt-days} .
\end{equation}
\vskip +0.5cm

\noindent
Kaspi et al. (2005) consider a radius-luminosity relation derived from the 
average lags of all the Balmer lines, but we prefer to focus on the relation 
based exclusively on \hbeta\ lags because the significant difference we find 
between the FWHMs of \halpha\ and \hbeta\ suggests that the lines form, at 
least in part, from kinematically and possibly spatially distinct regions.
The resulting relations are

\begin{eqnarray}
\mbh &=& (1.3 \pm 0.3) \times 10^6 \nonumber \\
     & & \hskip -0.5in \left(
\frac{\lha}{10^{42}~{\rm erg~s^{-1}}} \right)^{0.57 \pm 0.06}\left(
\frac{\fwha}{10^{3}~\mathrm{\kms}} \right)^{2.06 \pm 0.06} \msun .
\end{eqnarray}
\vskip +0.5cm

\noindent

\begin{eqnarray}
\mbh &=& (2.4 \pm 0.3) \times 10^6 \nonumber \\
     & & \hskip -0.3in \left(
\frac{\lhb}{10^{42}~{\rm erg~s^{-1}}} \right)^{0.59 \pm 0.06}~\left(
\frac{\fwhb}{10^{3}~\mathrm{\kms}} \right)^2 \msun .
\end{eqnarray}
\vskip +0.5cm

Kaspi \etal\ also derive a new \rblr-\lhb\ relation from the
reverberation-mapped sample.  Their slope of 0.69 $\pm$ 0.06 differs
somewhat from the 0.56 $\pm$ 0.02 found here (Eq. 7).  While the small
difference in adopted cosmology plays some role, most of the
discrepancy arises from the different luminosity ranges covered by our
respective samples.  Our \hbeta\ luminosities range from $\sim
10^{39.5}$ to $10^{43}$~erg s$^{-1}$, whereas Kaspi et al.'s sample
has very few points 
below $10^{41}$~erg s$^{-1}$.  When use a
restricted sample with \lhb\ $> 10^{41}$~erg s$^{-1}$, we obtain a
\lf-\lhb\ relation with a shallower slope of 0.97 $\pm 0.04$, which
would lead to a steeper \rblr-\lhb\ relation, consistent with that
found by Kaspi \etal.  A homogeneous sample covering as wide a
luminosity range as possible is required to address the issue of
whether the slope of the \rblr-\lhb\ relation is truly
luminosity-dependent.

\epsfig{file=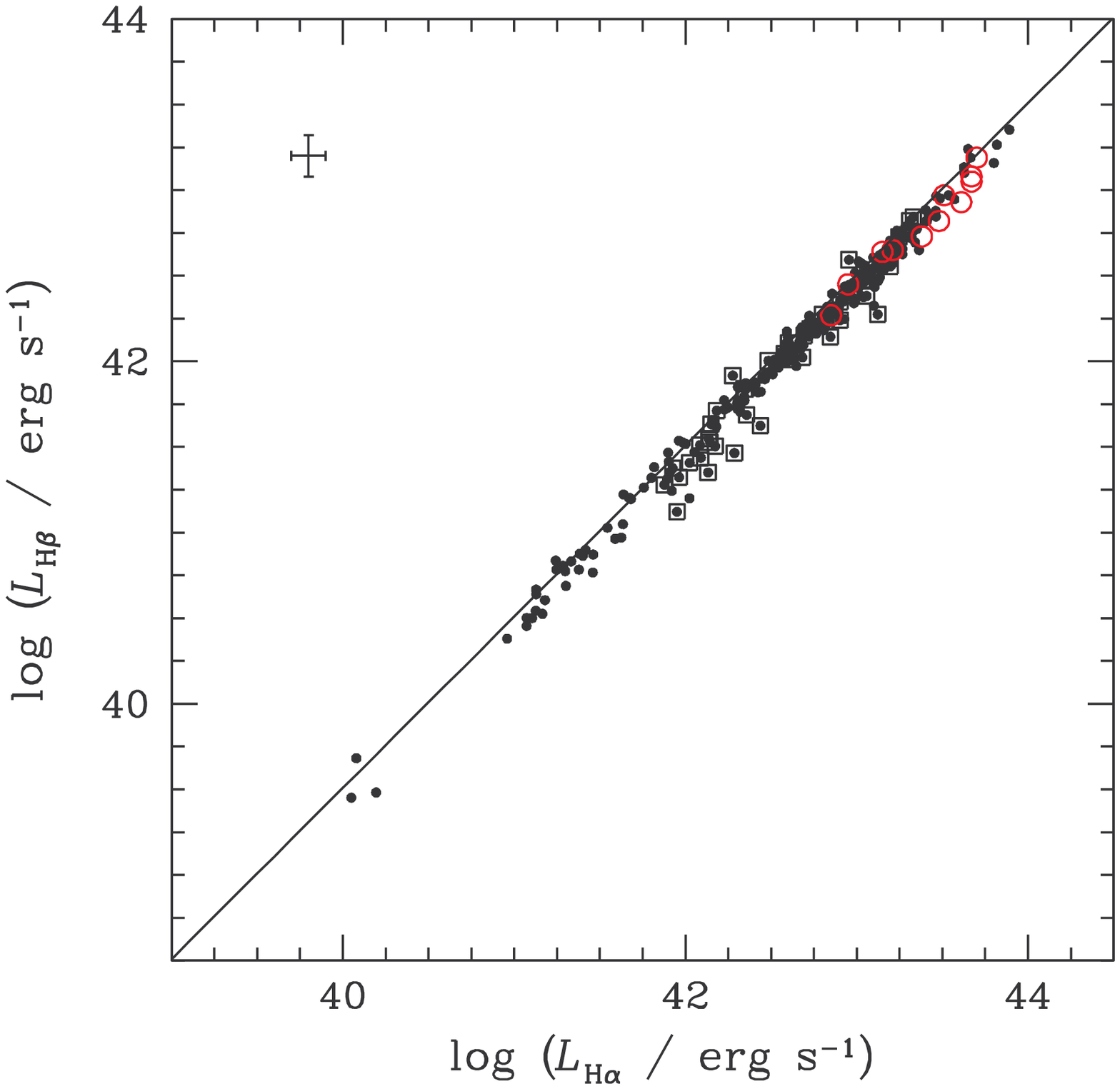,width=0.45\textwidth,keepaspectratio=true}
\vskip -1mm 
\figcaption[]{ 
Correlation between \lhb\ and \lha, for the combined broad
and narrow components of the lines.  Filled points are our
measurements, and open (red) points are from Kaspi \etal\ (2000);
radio-loud objects have boxes around them.  The line represents \lha\
= 3.1~\lhb, as expected for Case B\arcmin\ recombination.  The data
follow this relation quite closely.  A typical error bar is shown in
the upper-left corner.
\label{hahblum}}
\vskip +5mm

\section{Discussion}

The new virial mass formalism based on the \halpha\ emission line,
which is ultimately calibrated against the mass scale of
reverberation-mapped AGNs, relies on two strong empirical correlations
presented in this paper.  The first correlation, that between optical
continuum luminosity and Balmer emission-line luminosity, has long
been known (e.g., Yee \& Oke 1978; Yee 1980).  Shuder (1981) compiled
many of the early measurements, spanning a wide range of \halpha\
luminosity from $\sim 10^{40}$ to $10^{44}$~erg~s$^{-1}$.  He showed
that \lha\ traces the optical continuum luminosity with a slope of
$1.05 \pm 0.03$, consistent with unity and very similar to the slope
found here ($1.157 \pm 0.005$).  Similarly, from an analysis of AGNs
ranging from nearby low-luminosity Seyferts to quasars, Ho \& Peng
(2001) found that the luminosity of the broad \hbeta\ line, from $\sim
10^{38}$ to $10^{44}$~erg~s$^{-1}$, correlates strongly with the
$B$-band continuum luminosity, roughly of the form \lhb $\propto
L_B^{0.9}$.  While this slope is formally shallower than that found in
our sample ($1.133 \pm 0.005$), the measurements for the
low-luminosity sources in the sample of Ho \& Peng are also
considerably more uncertain, making it difficult to judge the
significance of the apparent discrepancy.  Our study, based on a large,
uniform, and modern data set, provides an important confirmation of
the earlier results.

The correlation between the luminosity of the Balmer lines and the
optical continuum arises naturally if the latter traces the low-energy
tail of a featureless continuum that extends into the far-ultraviolet,
which powers the line emission through photoionization.  Indeed, quite
early on this simple picture provided a natural explanation for the
observed uniformity of the EW of the broad \hbeta\ line in AGNs
(Searle \& Sargent 1968).  Recall that the slopes of the Balmer line
versus continuum luminosity relations (Eqs. 1 and 2) are formally
steeper than unity (1.16 and 1.13 for \halpha\ and \hbeta,
respectively).  This implies that the line strength increases mildly
with increasing luminosity [EW(\halpha)~$\propto$~\lf$^{0.16}$;
EW(\hbeta)~$\propto$~\lf$^{0.13}$], which means that the Balmer lines
show an {\it inverse}\ Baldwin effect (Baldwin 1977).  While various
studies have found a roughly linear relation between optical continuum
luminosity and \lhb\ (e.g.,~Yee 1980; Shuder 1981; Boroson \& Green
1992; Dietrich \etal\ 2002), Croom \etal\ (2002), consistent with our
study, report a positive slope of EW(\hbeta)~$\propto$~\lf$^{0.19}$
using composite spectra constructed from the 2dF+6dF quasar survey.
These apparently conflicting results may be attributable to the
different luminosity coverage in the different studies.  As discussed
by Croom et al., the luminosity-dependent variation of the Balmer line
strengths may indicate that the shape of the ionizing continuum
changes systematically with luminosity.  Another possibility is that
the covering factor of the line-emitting gas depends on luminosity.

The second major empirical correlation presented in this study links
the line widths of the \halpha\ and \hbeta\ lines.  While a number of
studies have investigated the relative differences in the profiles of
the \halpha\ and \hbeta\ lines (e.g., Osterbrock \& Shuder 1982;
Shuder 1982, 1984; Crenshaw 1986; Stirpe 1991; Kollatschny 2003), ours
considers by far the largest and most homogeneous sample to date.
Consistent with previous studies, we find that \hbeta\ generally has a
slightly broader profile than \halpha.  We measure
$\langle$\fwhb/\fwha$\rangle = 1.17$ with an error in the mean of 0.01
and an rms of 0.2.  This value is, in fact, identical to
the average value found by Osterbrock \& Shuder (1982), but
considering our order-of-magnitude increase in sample size this
agreement must be fortuitous.  The relative profile of \halpha\ and
\hbeta\ depends on both the kinematic and ionization structure of the
BLR, since \hbeta\ is emitted preferentially in regions of higher
density and/or higher ionization parameter than \halpha\
(e.g.~Osterbrock 1989).  The tendency for the \hbeta\ profile to be
broader than \halpha\ is not surprising if the density or ionization
parameter of the BLR increases with decreasing radii.

Further insights into the physical conditions of the BLR may be
gleaned from the Balmer decrements that come as a by-product of our
analysis.  The ratio of broad \halpha\ to \hbeta\ strengths has been
found to vary quite significantly from the value of 3.1 predicted by
Case B\arcmin\ recombination.  For instance, while Seyfert 1 galaxies
have typical \halpha/\hbeta\ ratios of $\sim 3.5-5$ (Adams \& Weedman
1975; Osterbrock 1977), in both broad-line radio galaxies and Seyfert
1.9 galaxies values as high as 7--10 are seen (Osterbrock 1981).
Interestingly, we find $\langle$\halpha/\hbeta$\rangle=3.5$ (for the
combined broad and narrow components; Fig. 5), which is virtually
identical to the value seen in the SDSS quasar composite spectrum of
Vanden Berk et al. (2001).  [The mean \halpha/\hbeta\ ratio for the
radio-loud group is slightly higher ($\sim$4), but given the smaller
sample, this is only a marginally significant difference.]  Since this
value is only mildly larger than the theoretical limit of 3.1, it
implies that processes that enhance the Balmer decrement, such as
collisional excitation or self-absorption (e.g., Netzer 1975), are
generally not that important.  We can also place a stringent limit on
the amount of internal reddening by dust: for an intrinsic Balmer
decrement of 3.1, the extinction law of Cardelli et al. (1989) would
allow at most $E(B-V)$ = 0.12 mag.  Most likely our selection of
AGN-dominated sources has biased us against objects with large Balmer
decrements, since the Balmer decrement may steepen in AGNs in the low
state (e.g.,~Tran, Osterbrock, \& Martel 1992; Korista \& Goad 2004).

We will revisit the empirical correlations discussed above in a
forthcoming study that considers a much larger and more
statistically complete sample.

\vskip +0.5cm

\section{Summary}

We present a new method for estimating black hole masses from
single-epoch optical spectra of AGNs, based entirely on the luminosity
and line width of the broad \halpha\ line.  We achieve formal
uncertainties comparable to those of the previous, widely used method
of Kaspi \etal\ (2000) that relies on the AGN optical continuum
luminosity and the width of the broad \hbeta\ line.  The advantage of
our method is that it enables us to derive robust masses in objects
where the AGN luminosity may be difficult (e.g., jet-dominated
systems) or impossible (e.g., low-luminosity AGNs with significant
galaxy contamination) to measure accurately, or in objects where the
broad \hbeta\ line may be missing altogether (e.g., Seyfert 1.9
galaxies).  The robustness of the \halpha-based mass estimator is due
to two tight empirical correlations we have established: that between
AGN optical continuum luminosity and Balmer emission-line luminosity
on the one hand, and that between \halpha\ and \hbeta\ line width on
the other.  We briefly discuss some physical ramifications of these
empirical correlations.

\acknowledgements 
We thank T.~Boroson for providing us with the {\sc
I}~Zw~1 iron template and an anonymous referee for thoughtful
comments that improved the manuscript.  L.~C.~H. acknowledges support
by the Carnegie Institution of Washington and by NASA grants from the
Space Telescope Science Institute (operated by AURA, Inc., under NASA
contract NAS5-26555).  We are grateful to the SDSS collaboration for
providing the extraordinary database and processing tools that made
this work possible.


\end{document}